\documentclass[twocolumn,showpacs,showkeys%
,amssymb,amsmath,nobibnotes,aps,prd]{revtex4}

\usepackage{amsmath}%
\usepackage{graphicx}%

\begin{document}

\title{Surface-Gravity Inequalities and Generic Conditions \\ for Strong Cosmic Censorship}
\author{Wenceslao \surname{Santiago-Germ\'{a}n}}
\affiliation{Department of Physics, University of Newcastle upon Tyne, NE1 7RU,U.K.}
\email{g.w.santiago-german@ncl.ac.uk}

\date{\today}

\begin{abstract}

Transforming Penrose's intuitive picture of a \textit{strong} cosmic
censorship principle\textemdash that generically forbids the appearance of locally naked space-time singularities\textemdash into a formal mathematical proof, remains at present, one of the most outstanding unsolved mathematical problems from the theory of gravitational collapse. Part of the difficulty lies in the fact that  we do not possess yet a clear-cut understanding of the hypothesis  needed for the establishment of some sort of strong cosmic censorship  theorem. What we have is a selected list of solutions, which at first sight seem to go against cosmic censorship, but at the end they fail in
some way. However, the space of solutions of Einstein's field equations is vast. In this article, we plan to increase one's intuition by
establishing a link between certain inequalities for
Cauchy-horizon stability and a set of generic conditions, such as 
a reasonable equation of state---which determines whether the space-time is asymptotically flat or not, an energy condition, and an
hypothesis over the class of metrics on which Einstein's field
equations ought to be solved to ensure strong cosmic censorship
inside black-holes. With  these tools in hand we examine the Cauchy-horizon stability of the theory created by Born and Infeld\textemdash whose 
action principle has been used as a prototype in superstring theory, and
the singularity-free Bardeen's black-hole model.  

\end{abstract}
\maketitle

\section*{\label{sec:intro}Introduction.}

One of the most striking predictions of Einstein's theory of
gravity is the occurrence of space-time singularities, i.e. causal
geodesic incompleteness,  as a generic feature of classical
gravitational collapse. This remarkable conclusion is
encapsulated by the singularity theorems of Hawking and Penrose,
who rely for its proof on techniques of differential topology
\cite{Penrose1970}. In principle, uncontrollable information may enter the space-time from a singularity and spoil one's ability to predict the future. That possibility has motived the following conjecture due to R. Penrose  \cite{Penrose1979}:

``Generically, the classical general theory of relativity would
not allow for the creation of  locally naked space-time
singularities."
 
This statement has been referred to as the \textit{strong cosmic censorship conjecture} and none knows whether it is true or false. 
 According to this viewpoint, the internal structure of a generic black-hole is bounded by a singularity that does not lie
into the past of any space-time point. 
 
 The proviso of generic is necessary since, for instance, the maximal analytical extension of the Kerr space-time, which describes an spinning black hole, offers a rich set of possibilities for experience including: close-time like curves, causal trajectories skirting the singularity, and new asymptotically
flat regions. However, this analytic extension can not survive
the propagation of small perturbations. These perturbations are
magnified without limit near the Cauchy horizon, which is then
transformed into a singular barrier that seals the region where
predictability is lost.
 
There are some mathematical results where one 
does not let\textemdash just by hand\textemdash the initial data develop or evolve outside some restriction on the metric, usually an
isometry.  The most complete study of this kind, up today, is the
one by Isenberg and Moncrief  on \textit{polarized Gowdy metrics}
\cite{IsenbergM}, which are vacuum (or electrovac) solutions of
Einstein's field equations having only two Killing vector fields.
Remarkably, the strong cosmic censorship conjecture was verified 
within this  restricted space of solutions.
In the Cauchy-horizon stability approach,  one must, in contrast,
examine  every conceivable perturbation to the space-time background
 which may lead to instability, only restricted by what
one should consider physically viable.
 
Before developing a general treatment that includes all the spinning 
black-holes, it is natural to look for clues in charge black-hole space-times  since they share certain resemblance with the causal structure of a rotating hole. 

Let us thus consider more carefully the Reissneer-Nordstr\"{o}m  solution. The  mathematical analysis on the propagation of small perturbations of \textit{compact support} in the Reissneer-Nordstr\"{o}m background leads to an \textit{unbounded} flux of radiation at the future-Cauchy horizon. The \textit{slowest rate of  divergence}\textemdash which takes the form
$e^{(\kappa_--\kappa_+)\upsilon},$ where the advanced time,
$\upsilon,$ goes to plus infinity  near the Cauchy horizon\textemdash is controlled by an inequality between the surface gravities, $\kappa_-$ and $\kappa_+,$ at the Cauchy and future event horizons respectively \cite{Chandra}.

\begin{figure}[t]
\includegraphics[width=3in]{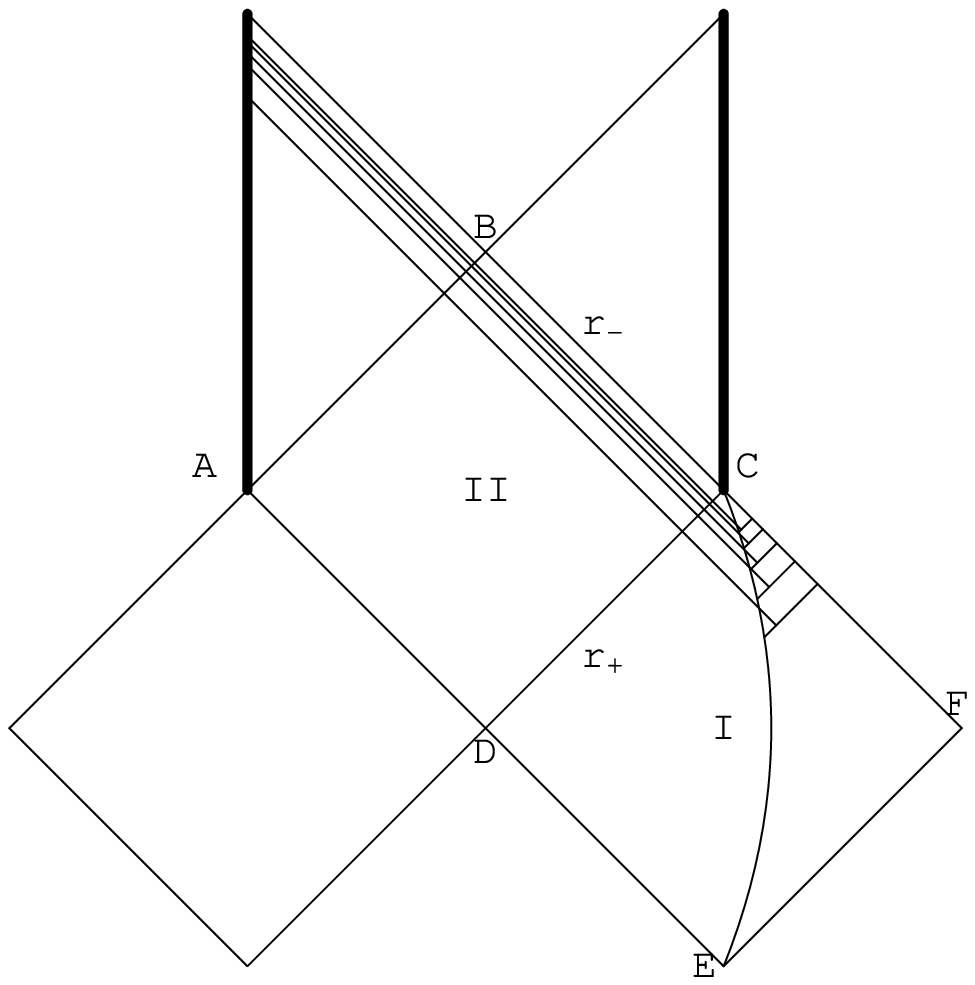}
\caption{\label{ChargeBH} Penrose diagram for a charge black-hole illustrating how the scattering of perturbations along the arc $\overline{EC}$ will result in  an accumulation of wave fronts  for the radiation received by an observer crossing the Cauchy horizon along the segment  $\overline{BC}$. When the inequality $-(\kappa_+-\kappa_-)>0$ between the surface gravities, $\kappa_-$ and $\kappa_+,$ of  the Cauchy horizon, $\overline{ABC},$ and future  event horizon, $\overline{ADC},$ is satisfied,  the flux of radiation diverges for all physically reasonable perturbations, even including those with  \textit{compact} support. This assertion still holds in the presence of a positive cosmological when the space-time is asymptotically De Sitter \protect\cite{BradyMossMayer}. The $\upsilon$\textendash coordinate  is defined to be and advanced time which takes the value $+\infty$ at $\overline{BC}$ and $-\infty$ at $\overline{AD}$.}
\end{figure}

That is,
\begin{equation}
 -(\kappa_+-\kappa_-)\geq 0. \label{Universal1}
\end{equation}
It seems, however, that the role of the \textit{surface-gravity inequality} (\ref{Universal1}), is of a more general character beyond the Reissner\textendash Nordstr\"{o}m geometry. Its relevance  arises because,
 heuristically, the radiation of compact support  is not only blue shifted
 by approaching the Cauchy horizon but also redshifted  by climbing away from the event horizon \cite{BradyMossMayer}, see FIG.\ref{ChargeBH}. Mathematically, the form of this divergence follows from the analytical properties of the
black hole scattering potential, or  potential well, which
vanishes at both horizons.

  We argue  that this universal inequality, (\ref{Universal1}), cannot be
 overcome, precisely, because it turns out to be equivalent \textemdash via  Einstein's field equations \textemdash to a \textit{local} condition  on the  source of the gravitational field, which is satisfied by all matter fields
obeying some \textit{energy condition} and a \textit{reasonable equation
of state} (which determines whether the space-time is asymptotically flat or not). Let us illustrate our assertion with an example.

\section{Spherical configurations of charge and mass}\label{round}
   To simplify things as much as possible, let us consider the metric for a \textit{static} spherically symmetric gravitational field produced by a
spherically symmetric body at rest, i.e.
\begin{eqnarray}
ds^2&=&-(1-\frac{2Gm(r)}{r}){s(r)}^2dv^2+2s(r)dvdr
\nonumber \\ &&
+r^2(d\theta^2+\sin^2\theta d\varphi^2), \label{line}
\end{eqnarray}
where $r$ denotes the area radius function. One may think of this
metric as roughly describing the asymptotically late-time behavior
of a nearly spherical  body undergoing gravitational collapse. To
simplify further, take a stress-energy tensor satisfying the equation
$T^{v}_{v}=T^{r}_{r}$ in Einstein's field equations. The result is that  $s(r)$ is now a constant function which can be normalized  to plus (or
minus) one. For the first option, the area radius function  is
decreasing towards the future along the ray $v=cte.$ 
 
 Theories of \textit{non-linear}  electrodynamics generate solutions of this kind and the black-holes they produce obey the zero and the first law of black-hole mechanics \cite{Rashdeed}. 
 
 In these coordinates, Einstein's law of gravitation leads to (See \cite{Husain})
\begin{equation}
 \rho=\frac{m_{,r}}{4\pi r^2 }, \label{density}
\end{equation}
\begin{equation}
2p=-\frac{m_{,rr}}{4\pi r } \label{pressure};
\end{equation}
where the mass function, $m(r),$ is given  by
\begin{eqnarray}
 \partial^\mu r \partial_\mu r &=& (1-\frac{2Gm(r)}{r})
\\ &\equiv& \frac{\Delta}{r^2},  \label{zeros}
\end{eqnarray}
and we have introduced the quantity $\Delta$
for notational purposes. The stress-energy tensor results to be diagonal and it has the form $(T^{v}_{v},T^{r}_{r},T^{\theta}_{\theta}, T^{\phi}_{\phi})=(-\rho,-\rho,p,p)$.  

If a static observer has four velocity
$(U^v,U^r,U^\theta,U^\phi)=(\Delta^{-1/2}r,0,0,0),$ he would
measure a local energy density, $\rho,$ and pressure, $p,$ equal to the
quantities  showed in equations (\ref{density}) and
(\ref{pressure}) respectively.

\section {The role of almost extreme black-holes}\label{ExtremeConditions}

Extreme black-holes have zero surface gravity, $\kappa=0$, and in
the charge-mass parameter space define a boundary beyond which
black-holes do not exist.  Therefore, they provide  an excellent
place to start an analysis about  cosmic censorship.

 Let us denoted by $\epsilon$ any set of parameters such as the mass, the
electric, or magnetic charges of some  classical field theory.
Although in general, given $\epsilon$, there are no algebraic
expressions relating the locations of the horizons of the black-holes  generated by these theories with the above physical parameters\textemdash for instance, in the  Born-Infeld theory, one needs to resort to elliptic integrals of the first kind\textemdash a simple rule applies for near  extreme black-holes. A rule, that will allow us to surmount certain technical difficulties and, in fact, avoid the direct calculation of the roots of the equation $\Delta=0$\textemdash which is generic\textemdash in order to evaluate the surface gravity inequality (\ref{Universal1}). 
 
 An extreme black-hole satisfy $\Delta=\Delta'=0$ at the
horizons, where the prime denotes the partial derivative with
respect to the area radius function\textemdash from there that its surface
gravity vanishes  since $\kappa=|\frac{\Delta'}{2r^2}|$. Taylor expand $\Delta$ around any of the extreme
values in r and $\epsilon$, say for definiteness $r_o$ and
$\epsilon_o,$ then
\begin{eqnarray}
\Delta(\epsilon,r)&=&
\Delta_{,\epsilon}\delta\epsilon+\frac{1}{2}\Delta_{,\epsilon \epsilon}\delta \epsilon^2
+\Delta'_{,\epsilon}\delta\epsilon\delta r
\nonumber \\& &
+\frac{1}{2} \Delta''{\delta r}^2+
 o\left ( (\delta r^2+\delta \epsilon^2)^{3/2} \right ).
 \end{eqnarray}
At second order, if $\Delta'' \neq 0 $, we end with a quadratic polynomial in $\delta r$. Take $\delta \epsilon$ in the direction that allows the existence of two real roots
\begin{equation}
\delta r= \pm \frac{1}{\Delta''}
 (-2\Delta'' \Delta_{,\epsilon}\delta\epsilon)^{1/2}
 -\frac{\Delta'_{,\epsilon}}{\Delta''}\delta\epsilon
+ o\left(\delta \epsilon^{3/2}\right).
\end{equation}
 Thus, it is possible to know the positions of the horizons and consequently evaluate the corresponding surface gravities.  The result is
\begin{eqnarray}
\kappa&=&|\pm \frac{1}{2r^2}(-2\Delta'' \Delta_{,\epsilon} \delta\epsilon)^{1/2}
\nonumber \\ & &
-\frac{\Delta_{, \epsilon}}{2\Delta'' r^{3}}(r{ \Delta}'''-4 \Delta'')\delta\epsilon
+ o\left(\delta \epsilon^{3/2}\right)| \label{kappa}
\end{eqnarray}
 Here, it is  assumed that $g_{\mu \nu} \in C^3$. To  have a bounded flux of the incident radiation at the future-Cauchy horizon, it is needed that
\begin{equation}
\kappa_+-\kappa_- >0, \label{mastercondition}
\end{equation}
but for nearly extreme black-holes, whose line element is of the form  (\ref{line}), with $s(r)=1$, Eq.(\ref{kappa}) implies that (\ref{mastercondition})  is equivalent to
\begin{eqnarray}
 r\Delta'''-4\Delta''>0, \label{criteria1}
\end{eqnarray}
or more succinctly, in terms of the pressure given by Eq.(\ref{pressure})
\begin{eqnarray}
  \frac{\delta p}{\delta r}>0,  \label{criteria2}
 \end{eqnarray}
which must be evaluated at extreme values.  
 
 Expressions (\ref{criteria1}) or (\ref{criteria2}) form a simple criterion for examining the disruption of predictability  in the interior of nearly extreme black-holes. For instance, it can be seen immediately that despite
all the effects produced by the addition of a positive \textit{
cosmological constant}: ray focusing, \textit{compact} achronal cross
sections, etc.  It does not help to enforce the Cauchy-horizon-stability requirement (\ref{mastercondition}) since it would give a vanishing contribution  to (\ref{criteria2})! This is a  consequence of the $C^3$\textendash smoothness on the metric tensor \cite{MossWSG}. We would like to remark here that in the presence of a positive cosmological constant, say $\Lambda$, besides the stability requirement (\ref{mastercondition}) 
it is also required to satisfy the inequality $(\kappa_{\Lambda}-\kappa_{-})>0.$ This extra condition  results from the fact of having causally aligned both the cosmological and future black-hole event horizons \cite{BradyMossMayer}.

 Let us now evoke the principle of \textit{local} conservation of material energy over a small region of the space-time. Taking the $r$\textendash component of
$T^{\nu}_{\mu;\nu},$  we have $\frac{\delta \rho}{\delta
r}=-\frac{2(\rho+p)}{r}$. But if $p=p(\rho)$, it follows that
\begin{equation}
\frac{\delta p}{\delta r}=-
\frac{2(\rho + p)}{r }\frac{\delta p}{\delta \rho}. \label{pressrad}
\end{equation}
Hence, by (\ref{criteria2}) and (\ref{pressrad}), both the \textit{weak
energy condition,} i.e. $\rho \geq 0$ and $p+\rho \geq0,$ and
an equation of state for the source of gravitational field
satisfying $\frac{\delta p}{\delta \rho}\geq 0$  imply the \textit{validity} of the strong cosmic censorship principle inside  black-holes closed enough to extremality, that possess a $C^3\text{--metric}$ of the form (\ref{line}) with $s(r)=cte$.

\begin{figure}[b] 
\includegraphics[width=2.5in]{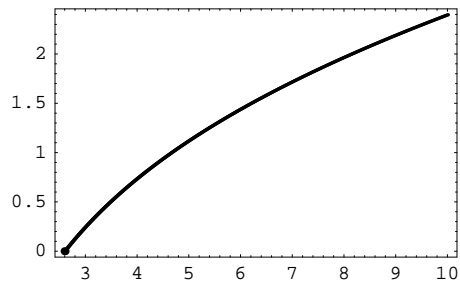} 
\caption{\label{BardeenBH} $-|g|(\kappa_+-\kappa_-)$ vs. $2GM/|g|.$ This graph shows the Cauchy-horizon instability of  Bardeen's black-hole.
Near the extreme value $2M_oG/|g_o|=\frac{3\sqrt{3}}{2}$ (small circle), the instability can be deduced from Eq.(\ref{p'Bar}). Whereas when  $M$ goes to infinity,  $-|g|(\kappa_+-\kappa_-)$ approaches to $(2GM/|g|)^{1/2}$ remaining positive.}
\end{figure}

\section{Bardeen black-holes}
The Bardeen model is a singularity-free black-hole solution representing the self-gravitating field of a non-linear magnetic monopole  satisfying the weak energy condition \cite{Bardeen}. The line-element of this hole has the form  (\ref{line}), with $s=1,$ but the mass-function is given by
\begin{equation}
m(r)=\frac{M r^3}{(r^2+g^2)^{3/2}},
\end{equation}
where $g$ denotes the monopole charge and  $M$ is the \textsc{ADM} mass. Black-hole space-times occur when the monopole charge is bounded by the \textsc{ADM} mass as follows  
\begin{equation}
\frac{3\sqrt{3}}{2}|g|\leq 2MG.
\end{equation}
The limiting case, when this inequality is saturated, corresponds to an extreme black-hole whose horizon's area is equal to $\mathcal{A}=8\pi g^2.$ 

The maximal analytic extension of these space-times is free of space-time singularities. However, as can be readily seen by the Cauchy-horizon-stability criteria (\ref{criteria2}), near the extreme case, the analytic extension is unstable since 
\begin{equation}
\label{p'Bar}
 (\frac{\delta p}{\delta r})_{(\epsilon_0,r_o)}=
 -\frac{5M}{54\pi\sqrt{6} g^4}\leq 0.  
\end{equation}

This conclusion about instability holds true for the rest of the space
of physical parameters, as shown in FIG.\ref{BardeenBH}
 
\section{Variation formulas for no extreme black-holes}
\label{quasitheorem}
 
In view of the success of establishing a link between an inequality\textemdash which involves two closed but \textit{separated} events in 
the space-time\textemdash and the \textit{local} properties of the matter field that generates the near extreme black-hole background; one may ask whether the current \textit{local} hypothesis cannot be extended to cover all other
cases as well, where near extreme conditions do not occur.  If
cosmic censorship is a principle of classical physics, it should
manifest itself as a consequence of the local  rules in which  the
laws of classical physics are formulated. Hence, it seems
reasonable to imagine that one may proceed to develop  a
topological argument,  in which the vanishing of the quantity
$\kappa_+-\kappa_-$ is forbidden unless one  reaches a point 
representing an extreme black-hole.  Then, by continuity, and the results of the last section, the neighborhood where $-(\kappa_+-\kappa_-)\geq0$ should necessarily cover the whole simple-connected space of physical parameters that allows the existence of two horizons. Let us see what are the consequences of adopting this kind of approach. 

 Firstly, since it is technically difficult to find roots for non-algebraic equations, let us change of variables, from the total charge and \textsc{ADM} mass $(q^2,M)$ to the horizon positions $(r_{-},r_{+})$. Then, by virtue of Einstein's field equations, the determinant of the
relevant Jacobian is given by
\begin{equation}
\det \frac{\partial(r_-,r_+)}{\partial(q^2,M)}=\frac{4\pi G^2}{r_+ r_- \kappa_+ \kappa_-}\int^{r_{+}}_{r_-}(\partial_{,q^2}\rho) r^2dr.
\end{equation}
 This transformation is invertible if the horizons are not degenerate and a change in the total energy density,  due to an increase in the charge, $ \delta q^2,$ enclosed between  horizons is not zero.
Such a description also leads to the variation formula
\begin{widetext}
\begin{equation}
-\delta(\kappa_+-\kappa_-)=-G(\frac{{\kappa_+}_{,r_+}}{r_+\kappa_+}+\frac{{\kappa_-}_{,r_-}}{r_-\kappa_-})\delta M + 2\pi (w_+[m_{,q^2},\kappa]+w_-[m_{,q^2},\kappa])\delta q^2, \label{BHMechanics}
\end{equation}
\end{widetext}
where we have defined
\begin{equation}
w[f,g]\equiv f'g-g'f ;  \qquad
f'=\frac{1}{\frac{1}{4G}\kappa}\partial_\mathcal{A}f,
\end{equation}
to denote the Wronskian between two functions of the area element,
$\mathcal{A}=4\pi r^2.$ The next step is to evaluate  the function
$\chi(M,q^2)=-\{\kappa_+(M,q^2)-\kappa_-(M,q^2)\}$ at a critical
point. Let us assume that the pressures $p_+$ and $p_-,$ at $r_+$ and $r_-$
respectively, are \textit{positive.} Then, the vanishing of the variation of
$-(\kappa_+-\kappa_-)$ with respect to the \textsc{ADM} mass,
$M,$ implies
\begin{equation}
 \frac{p_+}{p_-}>\frac{\kappa_+}{\kappa_-}.
\end{equation}
 Thus, we have that $-(\kappa_+-\kappa_-)>0$ at a critical point of $\chi$ (if this exits) since $p$ is assumed to be $r$\textendash decreasing in order to cover the near extreme case. To prove
(\ref{Universal1}), we shall add a further hypothesis:
\begin{equation}
\lim_{r \rightarrow o^+}\rho=\frac{a^2}{8\pi G }r^{-\gamma}; \qquad
  \gamma \geq 2.  \label{Tipler}
\end{equation}
If $\gamma=2$, take $a^2>1$. In effect, Eq.(\ref{Tipler}) implies
that $\lim_{r_- \rightarrow o^+}-(\kappa_+-\kappa_-)=+\infty.$ Being 
$\chi(M,q^2)$ a continuous function that does not vanish except at 
an extreme point. We conclude, by varying $M$ and taking different values of $q^2,$ that the surface-gravity inequality (\ref{Universal1}) holds. In such circumstances, strong cosmic censorship holds for not necessarily
extreme black-holes as well.

\section{The Born-Infeld action}

 The postulate of the constancy of the velocity of light  in relativistic mechanics means that there  is an upper limit to the velocity of massive particles moving   along  a  path of maximal \textit{length}.  In 1934,  Born and Infeld  suggested to put forward the classical foundations  of electrodynamics  on a similar basis by adding two postulates:

 1. The existence  of an \textit{absolute} constant, $1/b,$ placing an upper limit  to the magnitude of purely electric  fields, and 2. The Lagrangian of the theory to be the square root  of a determinant (see Ref.\cite{{Born-Infeld}}).
 
  The \textit{Born-Infeld action} then follows from the requirement of
having Maxwell theory as the limit when $b$ goes to zero. Its Lagrangian density reads

 \begin{equation}
 L=\frac{1}{b}\{ \sqrt{-g}-\sqrt{-\det(g_{\mu \nu} + bF_{\mu \nu}})\}  \label{BornInfeld1}
\end{equation}
or
\begin{equation}
 L=\frac{1}{b}\{ 1-\sqrt{1+b^2({B}^2-{E}^2)-b^4({E} \cdot {B})^2 } \}. \label{BornInfeld2}
\end{equation}

By Birkhoff's theorem, the spherically symmetric electrovac solution of
Einstein's field equations coupled to a Born-Infeld field is \textit{static}
and has the form
\begin{equation}
\label{NEmetric}
ds^2=-(1-\frac{2Gm}{r})dt^2+(1-\frac{2Gm}{r})^{-1}dr^2+r^2d\Omega^2,
\end{equation}
\cite{Gibbons}, where 
\begin{eqnarray}
m(r)&=&\frac{1}{b^2}\int^r_0 (\sqrt{(4\pi x^2)^2 + b^2(Q^2 +P^2)}
-4\pi x^2)dx \nonumber
 \\ && -(\frac{(Q^2+P^2)^{3/4}}{3\sqrt{\pi b}}\int_0^\infty\frac{1}{\sqrt{y^4+1}}-M).
\end{eqnarray}
 It depends on the electric-magnetic dual invariant combination, $Q^2+P^2,$ that we set equal to $q^2$, and the \textsc{ADM} mass, $M,$ which appears as an integration constant.
Here, the electromagnetic energy and pressure are given by
\begin{eqnarray}
\rho=-L+E \cdot \frac{\partial L}{\partial E}  =\frac{1}{b^2}(\sqrt{1+ \frac{ b^2q^2}{(4 \pi r^2)^2}}-1)
\end{eqnarray}
and
\begin{equation}
p=L-B \cdot \frac{\partial L}{\partial B}
=\frac{1}{b^2}(1-\frac{1}{\sqrt{1+ \frac{b^2q^2}{(4 \pi r^2)^2}}})
\end{equation}
respectively. Therefore, its equation of state is
\begin{equation}
p=\frac{\rho}{1+b^2\rho} \label{State}
\end{equation}
and the dominant energy condition, $-\rho\leq p \leq \rho,$ holds. 

The spherically symmetric stationary black-holes of the Born-Infeld theory, whose pure electric field\textemdash we recall\textemdash has bounded magnitude, do not provide counterexamples to strong cosmic censorship since: \\
\indent \;\;i) Their energy density and  pressures are positive, \\
\indent \;ii) Their equation of state, of the form
$p=\frac{\rho}{1+b^2\rho},$ \\
\indent $\; \;$ satisfies $\frac{\delta p}{\delta \rho} \geq 0,$ \\
\indent iii) $\lim_{r \rightarrow o^+}\rho=\frac{2G|q|/b}{8\pi G }r^{-2}$.

The fact that $2G|q|/b>1$ is a necessary  requirement for having
two horizons with $r_+\neq r_-.$  Since $\gamma=2,$ we see that the  
Born-Infield theory corresponds to a limiting case of Eq.(\ref{Tipler})

\section{Generalization capability}
The criterion expressed by the inequality (see (\ref{criteria1}))
 \begin{equation}
 \{-(r\Delta'''-4\Delta'')\}_{(r_o,\epsilon_o)}\geq0, \label{CHstab}
 \end{equation}
which would imply\textemdash if verified\textemdash the
establishment of some sort of strong cosmic censorship principle,
can be generalized in a form that does not rely on the spherical
assumption, and where the nature of having two non degenerate Killing
horizons is seen manifestly.

In effect, for a sufficiently nearly extreme black-hole let us use a chart, covering the region between the two horizons, in a manner of the
\textsc{ADM} 3+1 splitting
\begin{equation}
ds^2=\alpha^2
d\upsilon+h_{ab}(dx^a+\beta^ad\upsilon)(dx^b+\beta^bd\upsilon),
\end{equation}
where $\upsilon^\mu$ is a Killing vector field, and hence, all the
components of the metric have no dependence  on the
$\upsilon$\textendash coordinate. 

The norm of this vector is given by
\begin{equation}
\upsilon^\mu \upsilon_\mu=g_{\upsilon
\upsilon}=\alpha^2+\beta^a\beta_a
\end{equation}
and at the horizons is assumed to vanish. $\upsilon^\mu$ aligns\textemdash
we assume\textemdash with a null geodesic generator $\xi^\mu$ so that 
\begin{equation}
\xi^\nu \xi^{\mu}_{;\nu}=\kappa \xi^{\mu},
\end{equation}
where $\kappa$ is the surface gravity along the horizon. 

Now, let the horizon positions be given by constant values of a scalar 
function $\beta$, and  let us choose  $\beta^a$ so that it approaches smoothly to the direction given by $\partial_{\beta},$ in the vicinity of the horizons. 

From this we observe that
\begin{equation}
\kappa=\Gamma^{\upsilon}_{\upsilon \upsilon}=-\frac{g_{\upsilon \upsilon,\beta}}{2s},
\end{equation}
where it is assumed that $g^{\upsilon \beta}$ goes to a non zero
function $s^{-1}$ at the horizon.

 Hence\textemdash by inspection of the analysis presented in section \ref{ExtremeConditions},  we should
take $\frac{-1}{s}g_{\upsilon \upsilon}$ instead of $\Delta$
as a function of $\epsilon$ and $\beta,$ and arrive to the
conclusion that the criterion (\ref{CHstab}) must be
replaced by the Cauchy-horizon-stability requirement
\begin{equation}
-(\beta \frac{\partial^3}{\partial \beta^3}-4
\frac{\partial^2}{\partial \beta^2})\frac{\upsilon^{\mu}
\upsilon_{\mu}}{s}\geq 0 \text{ at extreme values.} \label{CHadm}
\end{equation}
 
 It would remain its transcription in terms of the local properties of the matter sources of the gravitational field. In view of its dependence on the first three derivatives of the metric tensor, this can be achieved using the
continuity formula
\begin{equation} \frac{1}{\alpha}\partial_v
\rho + J^b|_b = \pm \left(  S^{ab}K_{ab} +  \rho K \right)+
\rho|_b\frac{\beta^{b}}{\alpha}-2(\ln \alpha)|_bJ^b,
\label{Bernoulli}
\end{equation}
and  the Euler  equation \cite{York},
\begin{eqnarray}
\frac{1}{\alpha}\partial_tJ^a+S^{ab}|_b&=&\pm \left( 2 J^b
K^a_b+ KJ^a \right)+ \frac{1}{\alpha}J^a  \nonumber \\ && -(\ln
\alpha)|_b(S^{ab}+\rho h^{ab}).
\end{eqnarray}

Perhaps the  double-null formalism will prove to be 
more profitable in this goal. But the message we wanted to cross this
time is merely that this can be done. 

One further remark is in order. It is possible that by other means the condition $g \in C^3$ could be refined somewhat. However, as was pointed out in Ref.\cite{MossWSG}, the metric tensor  has to be at least  $C^1$.
If this assumption is dropped, one could construct counterexamples
with space-like shells of matter, where the extrinsic curvature
 of the shell has a jump from one  side  to the other; yet the weak energy
condition still holds and there is safety in crossing  the
black-hole Cauchy horizon.

 In conclusion, there is hope that the transcription of the surface-gravity inequality (\ref{Universal1}), or (\ref{CHadm}), in terms of the local properties of the matter sources of the gravitational field, would provide in more complicated situations with a reasonable set of conditions from which one may elucidate a theorem for strong cosmic censorship.
\begin{acknowledgments}
I am particularly grateful to Professor I.G Moss for various
important remarks and  discussions. This research  has been sponsored
by \textsc{CONACyT} of M\'exico under the grant\textendash $116020$.
\end{acknowledgments}


\begin{thebibliography}{99}

\bibitem{Penrose1970} S.W. Hawking and R. Penrose,  Proc. Roy. Soc. Lond.,   A314, 529 (1970).
\bibitem{Penrose1979} R. Penrose, in \textit{General Relativity, an Einstein Centenary Survey}, eds. S.W. Hawking and W. Israel, (Cambridge University Press, London, 1979); M. Simpson and R. Penrose,  Internal instability in a Reissner-Nordstr\"om black hole, Int. J. Theor. Phys., \textbf{7}, 183-97 (1973).
\bibitem{IsenbergM} J. Isenberg, V. Moncrief,  Ann. of Physics \textbf{199,} 84-122 (1990).
\bibitem{Chandra} S. Chandrasekhar and J.B. Hartle,  Proc. Roy. Soc. (London) \textbf{484} 301 (1982) ; S. Chandrasekhar, \textit{The Mathematical Theory of Black Holes} (Oxford University Press, New York, 1998).
\bibitem{BradyMossMayer}P. R. Brady, I.G. Moss and R. Myers, Phys. Rev. Lett. \textbf{80} 3432 (1998).
\bibitem{Rashdeed} D.A. Rashdeed, \textit{Non-linear Electrodynamics:} \textit{Zeroth and First Law of Black Holes Mechanics} (1997); hep-th/9702087.
\bibitem{Husain} V. Husian, Phys. Rev. D \textbf{53} (4), 1759-1762 (1995); M. Visser, \textit{Lorentzian Wormholes: from Einstein to Hawking.} (Springer-Verlag, New York, 1996).
\bibitem{MossWSG} W. Santiago-Germ\'{a}n,  PhD Thesis. \textit{Space-time Structure and Hidden Dimensions:} University of Newcastle upon Tyne (2003); W. Santiago-Germ\'{a}n  and I. G. Moss, Class. Quantum Grav. \textbf{18} 5097-5101, (2001).
\bibitem{Bardeen} J. Bardeen, presented at GR5. Tiflis, U.S.S.R., and published in the conference proceedings in the U.S.S.R (1968); E. Ay\'on-Beato and A. Garc\'ia, Phys.Lett. B \textbf{493} 149-152 (2000).
\bibitem{Born-Infeld} M. Born and L. Infeld, Proc. Roy. Soc. (London) A144, 425 (1934).
\bibitem{Gibbons} G.W. Gibbons and D.A. Rasheed,  Nuclear Physics B \textbf{454, } 185-206 (1995),.  
\bibitem{York} J.W. York \textit{ The Initial Value Problem and Dynamics,} Gravitational Radiation, (North Holland, 1982).

\end{thebibliography}
\end{document}